\documentclass[journal]{IEEEtran}

\usepackage{multirow}
\usepackage[table,svgnames]{xcolor}
\usepackage{amsmath,graphicx}
\usepackage{mdframed}
\usepackage{amssymb}
\usepackage[ruled,vlined]{algorithm2e}
\usepackage[noend]{algpseudocode}
\usepackage{graphicx}
\usepackage{epstopdf}
\usepackage{enumerate}
\usepackage{url}

\newtheorem{theo}{Theorem}[section]
\newtheorem{lem}{Lemma}

\newcommand{\norm}{\mathcal N} 

\newcommand{\X}{\bo X}

\newcommand{\I}{\bo I}

\newcommand{\dims}{s}
\newcommand{\dimb}{b}  
\newcommand{\dimr}{r}

\newcommand{\R}{\mathbb{R}}      
\newcommand{\Po}{\mathbb{P}_n}      
\newcommand{\Ps}{\mathbb{P}_{n,b}^{(k)}}      
\newcommand{\Pb}{\mathbb{P}_{n,b}^*}      
\newcommand{\G}{\mathbb{G}}      
\newcommand{\Gb}{\mathbb{G}_{n,b}^*}      

\newcommand{\bom}[1]{\boldsymbol{#1}}    
\newcommand{\bo}[1]{\mathbf{#1}}              

\newcommand{\x}{\bo x}

\newcommand{\n}{\bo n}

 \newcommand{\h}{\mathcal  H}
\newcommand{\q}{\mathcal  Q}
\newcommand{\f}{\mathcal  F} 

\newcommand{\z}{\bo z}

\newcommand{\beq}{\begin{equation}}
\newcommand{\eeq}{\end{equation}}
\newcommand{\bmat}{\begin{pmatrix}}
\newcommand{\emat}{\end{pmatrix}}
 \newcommand{\beqa}{\begin{eqnarray}}
\newcommand{\eeqa}{\end{eqnarray}}

\newcommand{\eps}{\varepsilon} 

\newcommand{\al}{\alpha} 
\newcommand{\sig}{\sigma}

\newcommand{\bth}{\bom \theta} 
  
\newcommand{\tho}{\hat\bth_n}      						
\newcommand{\ths}{\hat\bth_{n,b}^{(k)}}     	
\newcommand{\thb}{\hat\bth_{n,b}^*}      			
\newcommand{\thf}{\hat\bth_{n,b}^{R*}}      	
      
\newcommand{\mX}{\bom{\mathcal X}}

\usepackage{cite}

\ifCLASSINFOpdf
\else
\fi
\ifCLASSOPTIONcompsoc
  \usepackage[caption=false,font=normalsize,labelfont=sf,textfont=sf]{subfig}
\else
  \usepackage[caption=false,font=footnotesize]{subfig}
\fi
\begin{document}

\title{Robust, scalable and fast bootstrap method for analyzing large scale data}
\author{Shahab~Basiri, Esa~Ollila,~\IEEEmembership{Member,~IEEE,}
        and~Visa~Koivunen,~\IEEEmembership{Fellow,~IEEE}}

%
\maketitle
\begin{abstract}
In this paper we address the problem of performing statistical inference for large scale data sets i.e., Big Data. The volume and dimensionality of the data may be so high that it cannot be processed or stored in a single computing node. We propose a scalable, statistically robust and computationally efficient bootstrap method, compatible with distributed processing and storage systems. 
Bootstrap resamples are constructed  with smaller number of distinct data points on multiple disjoint subsets of data, similarly to the bag of little bootstrap method (BLB) \cite{BLB}. Then significant savings in computation is
achieved by avoiding the recomputation of the estimator for each bootstrap sample. Instead, a computationally efficient fixed-point estimation equation is analytically solved via a smart approximation following the Fast and Robust Bootstrap method (FRB) \cite{FRB08}.
Our proposed bootstrap method facilitates the use of highly robust statistical methods in analyzing large scale data sets. 
The favorable statistical properties of the method are established analytically.
Numerical examples demonstrate scalability, low complexity and robust statistical performance of the method in analyzing large data sets.
\end{abstract}


%
\IEEEpeerreviewmaketitle

\section{Introduction}
\label{sec:intro}
\IEEEPARstart{R}{ecent} 
advances in digital technology have led to a proliferation of large scale data sets. Examples include climate data, social networking, smart phone and health data, etc. 
Inferential statistical analysis of such large scale data sets is crucial in order to quantify statistical correctness of parameter estimates and testing hypothesis. 
However, the volume of the data has grown to an extent that cannot be effectively handled by traditional statistical analysis and inferential methods. Processing and storage of massive data sets becomes possible through parallel and distributed architectures.  Performing statistical inference on massive data sets using distributed and parallel platforms require fundamental changes in statistical methodology. Even estimation of a parameter of interest based on the entire massive data set can be prohibitively expensive. In addition, assigning estimates of uncertainty (error bars, confidence intervals, etc) to the point estimates is not computationally feasible using the conventional statistical inference methodology such as bootstrap \cite{Ef}. 

The bootstrap method is known as a consistent method of assigning estimates of uncertainty (e.g., standard deviation, confidence intervals, etc.) to statistical estimates \cite{Ef, Dav} and it is commonly applied in the field of signal processing \cite{Zoubir, Zoubir2}. However, for at least two obvious reasons the method is computationally impractical for analysis of modern high volume and high-dimensional data sets:
\emph{First}, the size of each bootstrap sample is the same as the original big data set (with about  63\% of data points appearing at least once in each sample typically), thus leading to processing and storage problems even in advanced computing systems.  
\emph{Second}, (re)computation of value of the estimator for each massive bootstrapped data set is not feasible even for estimators with moderate level of computational complexity.
Variants such as subsampling \cite{sub} and the $m$ out of $n$ bootstrap \cite{mon} were proposed to reduce the computational cost of bootstrap by computation of the point estimates on smaller subsamples of the original data set. Implementation of such methods is even more problematic as the output is sensitive to the size of the subsamples $m$. In addition extra analytical effort is needed in order to re-scale the output to the right size. 

The bag of little bootstraps (BLB) \cite{BLB} modifies the conventional bootstrap to make it applicable for massive data sets. In BLB method the massive data is subdivided randomly into disjoint subsets (i.e., so called subsample modules or bags).
This allows the massive data sets to be stored in distributed fashion. Moreover subsample modules can be processed in parallel using distributed computing architectures. The BLB samples are constructed by assigning random weights from multinomial distribution to the data points of a disjoint subsample.
Although in BLB the problem of handling and processing massive bootstrap samples is alleviated, yet (re)computation of the estimates for a large number of bootstrap samples is prohibitively expensive. Thus, on the one hand BLB is impractical for many commonly used modern estimators that typically have a high complexity. Such estimators often require solving demanding optimization problems numerically. On the other hand, using the primitive LS estimator in the original BLB scheme does not provide a statistically robust bootstrap procedure as the LS estimator is known to be very sensitive in the face of outliers. 

In this paper we address the problem of bootstrapping massive data sets by introducing a low complexity and robust bootstrap method. The new method possesses similar scalability property as the BLB scheme with significantly lower computational complexity. Low complexity is achieved by 
utilizing for each subset a fast fixed-point estimation technique stemming from Fast and Robust Bootstrap (FRB) method \cite{FRB00, FRB02, FRB08}. It avoids (re)computation of fixed-point equations for each bootstrap sample via a smart approximation. Although the FRB method possesses a lower complexity in comparison with the conventional bootstrap, the original FRB is incompatible with distributed processing and storage platforms and it is not suitable for bootstrap analysis of massive data sets.
Our proposed bootstrap method is scalable and compatible with distributed computing architectures and storage systems, robust to outliers and consistently provides accurate results in a much faster rate than the original BLB method.
We note that some preliminary results of the proposed approach were presented in the conference paper \cite{own2}.

The paper is organized as follows. In Section \ref{sec:RBM}, the BLB and FRB methods are reviewed. The new bootstrap scheme (BLFRB) is proposed in Section \ref{sec:BLFRB}, followed by implementation of the method for MM-estimator of regression \cite{MMest}. In Section \ref{sec:SP} Consistency and statistical robustness of the new method are discussed. Section \ref{sec:exp} provides simulation studies and an example of using the new method for analysis of a real world big data set. Section \ref{sec:conc} concludes.
\section{Related bootstrap methods} \label{sec:RBM}
In this section, we briefly describe the ideas of the BLB \cite{BLB} and FRB \cite{FRB08} methods. The pros and cons of both methods are discussed as well.
\subsection{Bag of Little Bootstraps} \label{ssec:BLBmethod}

Let $\X=(\x_1 \ \cdots \ \x_n) \in \R^{d \times n}$ be a $d$ dimensional observed data set of size $n$. The volume and dimensionality of the data may be so high that it cannot be processed or stored in a single node. Consider $\hat{\bth}_n \in \mathbb{R}^d$ as an estimator of a parameter of interest $\bth \in \mathbb{R}^d$ based on $\X$. Computation of estimate of uncertainty $\hat{\bom\xi}^*$ (e.g., confidence intervals, standard deviation, etc,) for $\hat{\bth}_n$ is of great interest as for large data sets confidence intervals are often more informative than plain point estimates.

The bag of little bootstraps (BLB) \cite{BLB} is a scalable bootstrap scheme that draws disjoint subsamples $\check\X = (\check\x_{1} \ \cdots \  \check\x_{\dimb})\in \R^{d \times b}$ (which form "bags" or "modules") of smaller size $\dimb = \{\left\lfloor n^\gamma\right\rfloor| \gamma\in\left[0.6,0.9\right]\}$ by randomly resampling \emph{without} replacement from columns of $\X$. For example if $n=10^7$ and $\gamma=0.6$, then $b=15849$.
 For each subsample module, bootstrap samples, $\X^*$, are generated by assigning a random weight vector $\n^{*} = (n_{1}, \ldots, n_{b}^*)$ from $Multinomial(n, (1/\dimb)\textbf{1}_{\dimb})$ to data points of the subsample, where the weights sum to $n$. 
The desired estimate of uncertainty $\hat{\bom\xi}^*$ is computed based on the population $\hat\bth^*_n$ within each subsample module and the final estimate is obtained by averaging $\hat{\bom\xi}^*$'s over the modules.

In the BLB scheme each bootstrap sample contains at most $b$ distinct data points. 
Thus the BLB approach produces the bootstrap replicas with reduced effort in comparison to conventional bootstrap \cite{Ef}. Furthermore, the computation for each subsample can be done in parallel by different computing nodes. Nevertheless, (re)computing the value of estimator for each bootstrap sample for example thousands of  times is still computationally impractical even for estimators of moderate level of complexity. This includes a wide range of modern estimators that are solutions to optimization problems such as maximum likelihood methods or highly robust estimators of linear regression. The BLB method was originally introduced with the primitive LS estimator. Such combination does not provide a statistically robust bootstrap procedure as the LS estimator is known to be very sensitive in the face of outliers. Later in section \ref{sec:SP} of this paper we show that even one outlying data point is sufficient to break down the BLB results.  

\subsection{Fast and Robust Bootstrap} \label{ssec:FRBmethod}

The fast and robust bootstrap method \cite{FRB08,FRB00,FRB02} is  computationally efficient and robust to outliers in comparison with conventional bootstrap. 
It is  applicable for estimators $\hat{\bth}_n \in \mathbb{R}^d$ that can be expressed as a solution to a system of smooth fixed-point (FP) equations: \beq\hat{\bth}_n = \q (\hat{\bth}_n; \X),\label{frb1}\eeq where $\q : \mathbb{R}^d \rightarrow \mathbb{R}^d$. 
The bootstrap replicated estimator $\hat{\bth}_n^*$ then solves \beq\hat{\bth}_n^* = \q(\hat{\bth}_n^*; \X^*),\label{frb2}\eeq where the function $\q$ is same as in \eqref{frb1} but now  dependent on the bootstrap sample $\X^*$. Then, instead of  computing $\hat{\bth}_n^*$ from \eqref{frb2}, we compute: 
\beq
\hat{\bth}_n^{1*} = \q(\hat{\bth}_n; \X^*),
\label{eq:frb1step}
\eeq
where the notation $\hat{\bth}_n^{1*}$ denotes an approximation of $\hat{\bth}_n^*$ in \eqref{frb2} with initial value $\hat{\bth}_n$ based on bootstrap sample $\X^*$. In fact $\hat{\bth}_n^{1*}$ is a one-step improvement of the initial estimate.
In conventional bootstrap, one uses the distribution of $\hat{\bth}_n^*$ to estimate the sampling distribution of $\hat{\bth}_n$. Since the distribution of the one-step estimator $\hat{\bth}_n^{1*}$  does not accurately reflect the sampling variability of $\hat \bth$, but  typically underestimates it, a linear correction needs to be applied as follows:
\beq
\hat \bth_n^{R*} = \hat{\bth}_n + \big[\textbf{I} - \nabla \q(\hat{\bth}_n;\X)\big]^{-1}\big(\hat{\bth}^{1*}_n - \hat{\bth}_n\big),
\label{eq:frb}
\eeq 
where $\nabla \q\left(\cdot\right) \in \mathbb{R}^{d \times d}$ is the matrix of partial derivatives w.r.t. $\hat{\bth}_n$. Then under sufficient regularity conditions, $\hat{\bth}_n^{R*}$ will be estimating the limiting distribution of $\hat{\bth}_n$. In most applications, $\hat{\bth}_n^{R*}$ is not only significantly faster to compute than $\hat{\bth}_n^{*}$, but numerically more stable and statistically robust as well. However, the original FRB is not scalable or compatible with distributed storage and processing systems. Hence, it
is not suited for bootstrap analysis of massive data sets. The method has been applied to many complex fixed-point estimators such as FastICA estimator \cite{own}, PCA and highly robust estimators of linear regression \cite{FRB08}. 

\section{Fast and Robust Bootstrap for Big Data}\label{sec:BLFRB}

\begin{figure*}[!t]
  \includegraphics[width=18cm,height=8cm]{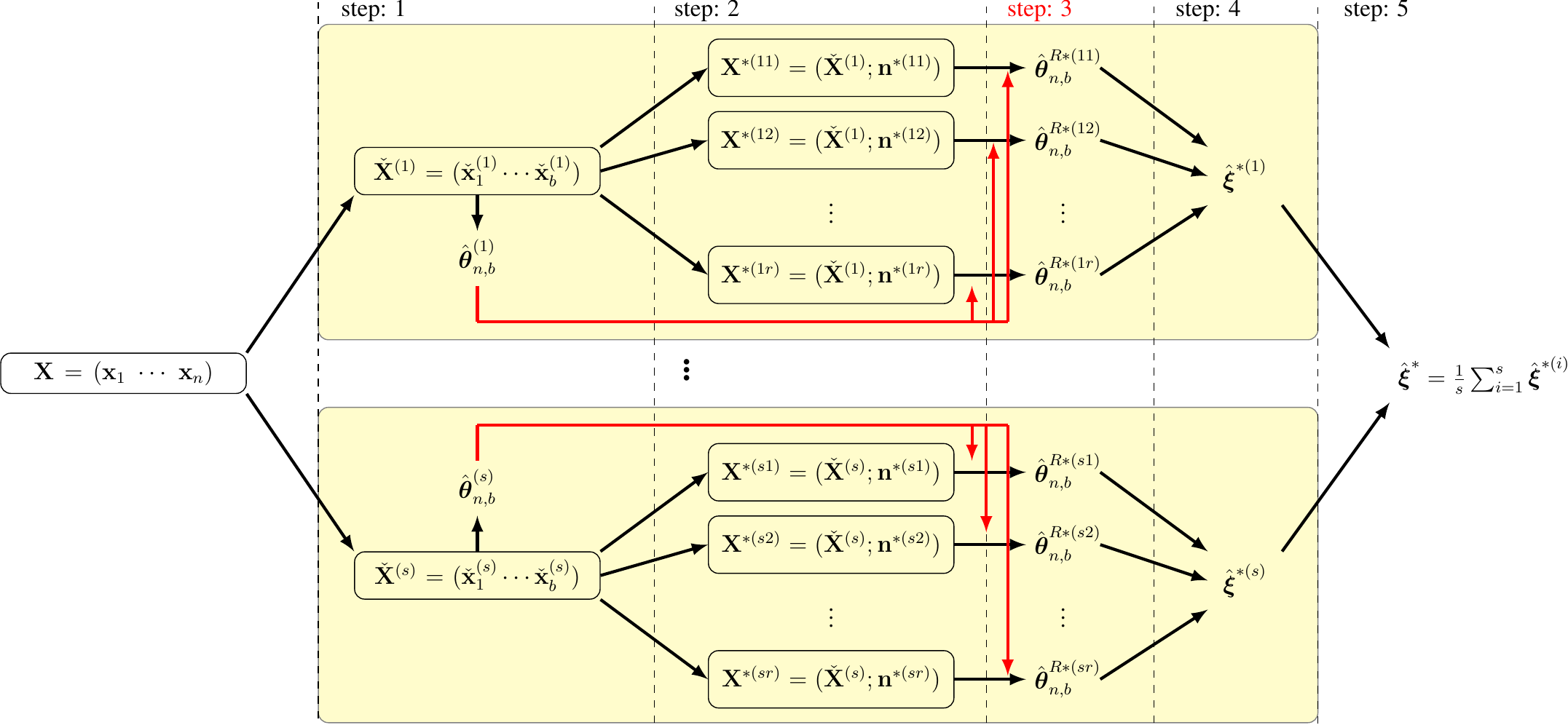}
	\caption{The steps of the BLFRB procedure (\textbf{Algorithm \ref{alg:BLFRB}}) are depicted. Disjoint subsamples of significantly smaller size $b$ are drawn from the original Big
Data set $\X$. The initial estimate $\hat\bth_{n,b}$ is obtained by solving fixed-point estimating equation only once for each subsample $\check\X$. Within each module, the FRB replicas $\hat\bth_{n,b}^{R*}$ are computed for each bootstrap sample $\X^*$ using the initial estimate $\hat\bth_{n,b}$. The final estimate of uncertainty $\hat{\bom\xi}^*$ is obtained by averaging the results of distinct subsample modules.} 
	\label{blb}
\end{figure*}

In this section we propose a new bootstrap method that combines the desirable properties of the BLB and FRB methods. The method can be applied to any estimator representable as smooth FP equations. The developed Bag of Little Fast and Robust Bootstraps (BLFRB) method is suitable for big data analysis because of its scalability and low computational complexity. Recall that the main computational burden of the BLB scheme is in recomputation of estimating equation \eqref{frb2} for each bootstrap sample $\X^*$. Such computational complexity can be drastically reduced by computing the FRB replications  as in \eqref{eq:frb} instead. This can to be done locally within each bag. 
Let $\hat{\bth}_{n,b}$ be a solution to equation \eqref{frb1} for subsample $\check\X \in \R^{d\times b}$: 
\beq
\hat{\bth}_{n,b} = \q (\hat{\bth}_{n,b}; \check\X).
\label{eq:blfrb1}
\eeq
Let $\X^* \in \R^{d\times n}$ be a bootstrap sample of size $n$ randomly resampled with replacement from disjoint subset $\check\X$ of size $b$; or equivalently generated by assigning a random weight vector $\n^{*} = (n_{1}, \ldots, n_{b}^*)$ from $Multinomial(n, (1/\dimb)\textbf{1}_{\dimb})$ to data points of $\check\X$. The FRB replication of $\hat{\bth}_{n,b}$ can be obtained by 
\beq
\hat \bth_{n,b}^{R*} = \hat{\bth}_{n,b} + \big[\textbf{I} - \nabla \q(\hat{\bth}_{n,b};\check\X)\big]^{-1}\big(\hat{\bth}^{1*}_{n,b} - \hat{\bth}_{n,b}\big),
\label{eq:blfrb}
\eeq 
where $\hat{\bth}_{n,b}^{1*} = \q(\hat{\bth}_{n,b}; \X^*)$ is the one-step estimator and $\nabla \q\left(\cdot\right) \in \mathbb{R}^{d \times d}$ is the matrix of partial derivatives w.r.t. $\hat{\bth}_{n,b}$.
The proposed BLFRB procedure is given in detail in \textbf{Algorithm \ref{alg:BLFRB}}. The steps of the algorithm are illustrated in \textbf{Fig. \ref{blb}}, where $\check\X^{(k)}$, $k = 1, \ldots, s$ denotes the disjoint subsamples and $\X^{*(kj)}$ corresponds to the $j$th bootstrap sample generated from the distinct subsample $k$. 
Note that the terms $\hat{\bth}_{n,b}$ and $\big[\textbf{I} - \nabla \q(\hat{\bth}_{n,b};\check\X)\big]^{-1}$ are computed only once for each bag.  

While the BLFRB procedure inherits the scalability of BLB, it is radically faster to compute, since 
the replication $\hat\bth^{R*}_{n,b}$ can be computed  in closed-form with small number of distinct data points. 
Low complexity of the BLFRB scheme allows for fast and scalable computation of confidence intervals for commonly used modern fixed-point estimators such as FastICA estimator \cite{own}, PCA and highly robust estimators of linear regression \cite{FRB08}.
\begin{algorithm}[!t]
\nl Draw $\dims$ subsamples (which form "bags" or "modules") $\check\X = (\check\x_{1} \ \cdots \  \check\x_{\dimb})$ of smaller size $\dimb = \{\left\lfloor n^\gamma\right\rfloor| \gamma\in\left[0.6,0.9\right]\}$ by randomly sampling \emph{without} replacement from columns of $\X$\;

\For {each subsample  $\check\X$}{
\nl  Generate $\dimr$  bootstrap samples by resampling as follows:  \emph{Bootstrap sample}  $\X^* =(\check\X; \n^*)$ 
is formed by assigning a random weight vector $\n^{*} = (n_{1}, \ldots, n_{b}^*)$ from $Multinomial(n, (1/\dimb)\textbf{1}_{\dimb})$ to columns of $\check\X$\;
\nl  Find the initial estimate $\hat\bth_{n,b}$ that solves \eqref{eq:blfrb1} and for each bootstrap sample $\X^*$ compute $\hat{\bth}_{n,b}^{R*}$ from equation \eqref{eq:blfrb}\;
\nl  Compute the desired estimate of uncertainty $\hat{\bom\xi}^*$ based on the population of $\dimr$ FRB replicated values $\hat{\bth}_{n,b}^{R*}$\;
}
\nl Average the computed values of the estimate of uncertainty over the subsamples, i.e., $\hat{\bom\xi}^* = \frac{1}{s}\sum_{k=1}^s \hat{\bom\xi}^{*(k)}$.
\caption{The BLFRB procedure}\label{alg:BLFRB}
\end{algorithm}


\subsection{BLFRB for MM-estimator of linear regression} \label{sec:MMest}

Here we present a practical example formulation of the method, where the proposed BLBFR method is used for linear regression. In order to construct a statistically robust bootstrap method, MM-estimator that lends itself to fixed point estimation equations is employed for bootstrap replicas. Let $\X = \{(y_1,\z_1^\top)^\top, \ldots, (y_n,\z_n^\top)^\top\}$, $\z_i \in \R^p$, be a sample of independent random vectors that follow the linear model:
\beq
y_i = \z_i^\top\bth + \sig_0 e_i \quad \text{for} \quad i = 1, \ldots, n,
\label{lrm}
\eeq
where $\bth \in \R^p$ is the unknown parameter vector. Noise terms $e_i$'s are i.i.d. random variables from a symmetric distribution with unit scale. 

Highly robust MM-estimators \cite{MMest} are based on two loss functions $\rho_0 : \R \rightarrow \R^+$ and $\rho_1 : \R\rightarrow \R^+$ which determine the breakdown point and  efficiency of the estimator,  respectively. The $\rho_0(\cdot)$ and $\rho_1(\cdot)$ functions are symmetric, twice continuously differentiable with $\rho(0) = 0$, strictly increasing on $[0, c]$ and constant on $[c, \infty)$ for some constant $c$.
The MM-estimate of $\hat\bth_n$ satisfies
\beq
\frac{1}{n}\sum_{i=1}^n \rho'_1\bigg(\frac{y_i - \z_i^\top \hat\bth_n}{\hat\sig_n}\bigg)\z_i = \textbf{0}
\label{eq3}
\eeq
where $\hat\sig_n$ is a S-estimate \cite{FRB84} of scale. Consider M-estimate of scale $\hat{s}_n(\bth)$ defined as a solution to
\beq
\frac{1}{n}\sum_{i=1}^n \rho_0\bigg(\frac{y_i - \z_i^\top \bth}{\hat{s}_n(\bth)}\bigg) = m,
\label{Sest}
\eeq
where $m=\rho_0(\infty)/2$ is a constant. Let $\tilde{\bth}_n$ be the argument that minimizes $\hat{s}_n(\bth)$,
\[
\tilde{\bth}_n = \arg \min_{\bth \in \R^p} \hat{s}_n(\bth),
\]
then $\hat\sig_n = \hat{s}_n(\tilde{\bth}_n)$.

We employ the Tukey's loss function: 
\[
\rho_e(u) = 
\begin{cases}
\frac{u^2}{2}-\frac{u^4}{2c_e^2}+\frac{u^6}{6c_e^4} & \text{for } |u| \leq c_e\\
\frac{c_e^2}{6} & \text{for } |u| > c_e
\end{cases},
\] 
which is widely used as the $\rho$ functions of the MM-estimator, where subscript $e$ represents different tunings of the function. For instance an MM-estimator with efficiency $\mathcal{O} = 95\%$ and breakdown point $BP = 50\%$ (i.e. for Gaussian errors) is achievable by tuning $\rho_e(u)$ into $c_0 = 1,547$ and $c_1 = 4,685$ for $\rho_0$ and $\rho_1$ respectively (see \cite[p.142, tab.19]{FRB87}).
In this paper \eqref{eq3} is computed using an iterative algorithm proposed in \cite{MMAlg}.
The initial values of iteration are obtained from \eqref{Sest} which in turns are computed using the FastS algorithm \cite{FastS}. 

In order to apply the BLFRB method to MM-estimator, \eqref{eq3} and \eqref{Sest} need to be presented in form of FP equations scalable to number of distinct data points in the data. 
The corresponding scalable one-step MM-estimates $\hat\bth_n^{1*}$ and $\hat\sig_n^{1*}$ are obtained by modifying \cite[eq. 17 and 18]{FRB08} as follows. 
Let $\X^*=(\check\X; \n^*)$ denote a BLB bootstrap sample  based on subsample $\check\X = \{(\check{y}_1,\check\z_1^\top)^\top, \ldots, (\check{y}_b,\check\z_b^\top)^\top\}$, $\check\z_i \in \R^p$ and a weight vector $\n^* = (n_{1}^* \cdots n_{b}^*) \in \R^b$,
\begin{gather}
\hat\bth_n^{1*} = \bigg(\sum_{i=1}^b n_i^*\check\omega_i \check\z_i \check\z_i^{\top} \bigg)^{-1}\sum_{i=1}^b n_i^*\check\omega_i \check\z_i \check{y}_i,
\label{eq5}
\\
\hat\sig_n^{1*} = \sum_{i=1}^b n_i^*\check\upsilon_i(\check{y}_i - \check\z_i^{\top} \tilde\bth_n),
\label{eq6}
\end{gather}
where 
\begin{gather*}
\check{r}_i = \check{y}_i - \check\z_i^\top \hat\bth_n, \qquad \tilde{r}_i = \check{y}_i - \check\z_i^\top \tilde\bth_n, \\
\check\omega_i = \rho'_1(\check{r}_i/\hat\sig_n)/\check{r}_i \quad\text{and}\quad \check\upsilon_i = \frac{\hat\sig_n}{nm} \rho_0(\tilde{r}_i/\hat\sig_n)/\tilde{r}_i.
\end{gather*}
The BLFRB replications of $\hat\bth_n$, are obtained by applying the FRB linear correction as in \cite[eq. 20]{FRB08}, to the one-step estimators of \eqref{eq5} and \eqref{eq6}.
\section{Statistical properties} \label{sec:SP}
Next we establish the asymptotic convergence and statistical robustness of the proposed BLFRB method.
\subsection{Statistical Convergence} \label{subsec:Consis}
We show that the asymptotic distribution of BLFRB replicas in each bag is the same as the conventional bootstrap.
Let $\X = \{\x_1, \ldots, \x_n\}$ be a set of observed data as the outcome of i.i.d. random variables $\mX = \{\mX_1, \ldots, \mX_n\}$ from an unknown distribution $P$. 
The empirical distribution (measure) formed by $\X$ is denoted by linear combination of the Dirac measures at the observations $\Po = n^{-1}\sum_{i = 1}^n \delta_{\x_i}$.
Let $\Ps = n^{-1}\sum_{i = 1}^b \frac{n}{b}\delta_{\check\x_i^{(k)}}$ and $\Pb = n^{-1}\sum_{i = 1}^n \delta_{\x_i^*}$ denote the empirical distributions formed by subsample $\check\X^{(k)}$ and bootstrap sample $\X^*$ respectively.
We also use $\phi(\cdot)$ for functional representations of the estimator e.g., $\bth = \phi(P), \ths = \phi(\Ps)$ and $\thb = \phi(\Pb)$. The notation $\stackrel{d}{=}$ denotes that both sides have the same limiting distribution.
\begin{theo} Consider $P$, $\Po$ and $\Ps$ as maps from a Donsker class $\f$ to $\R$ such that $\f_\delta = \{f-g: f,g \in \f, \{P(f - Pf)^2\}^{1/2} < \delta\}$ is measurable for every $\delta > 0$.
Let $\phi$ to be Hadamard differentiable at $P$ tangentially to some subspace and $\tho$ be a solution to a system of smooth FP equations. Then as $n,b \rightarrow \infty$
\beq\sqrt{n}(\thf - \ths) \stackrel{d}{=} \sqrt{n}(\tho - \bth).\label{eq:cons}\eeq See the proof in the Appendix.
\label{cons}
\end{theo}
\subsection{Statistical robustness}\label{subsec:MMest}
Consider the linear model \eqref{lrm} and let $\hat\bth_n$ be an estimator of the parameter vector $\bth$ based on $\X$. Let $q_{t}, t \in (0,1)$, denote the $t$th upper quantile of $[\hat\bth_n]_l$, where $[\hat\bth_n]_l$ is the $l$th element of $\hat\bth_n$, $l=1,\ldots,p$. In other words $Pr\big([\hat\bth_n]_l > q_{t} \big) = t$. Here we study the robustness properties of BLB and BLFRB estimates of $q_{t}$. 
We only focus on the robustness properties of one bag as it is easy to see that the end results of both methods break down, if only one bag produces a corrupted estimate. 

Let $\hat{q}_t^{*}$ denote the BLB or BLFRB estimate of the $q_t$ based on  a random subsample $\check\X$ of size $\dimb = \{\left\lfloor n^\gamma\right\rfloor| \gamma\in\left[0.6,0.9\right]\}$ drawn from a big data set $\X$.   
Following \cite{Singh98}, we define the upper breakdown point of $\hat{q}_t^{*}$ as the minimum proportion of asymmetric outlier contamination in subsample $\check\X$ that can drive $\hat{q}_t^{*}$ over any finite bound.

\begin{theo}
In the original BLB setting with LS estimator, only one outlying data point in a subsample $\check\X$ is sufficient to drive $\hat{q}_t^{*}$, $t \in (0,1)$ over any finite bound and hence, ruining the end result of the whole scheme. See the proof in the Appendix.
\label{blbrob}
\end{theo}

Let $\X = \{(y_1,\z_1^\top)^\top, \ldots, (y_n,\z_n^\top)^\top\}$, be an observed data set following the linear model \eqref{lrm}. Assume that the explanatory variables $\z_i \in \R^p$ are in general position \cite[p. 117]{FRB87}. Let $\hat\bth_n$ be an MM-estimate of $\bth$ based on $\X$. According to \cite[Theorem 2]{FRB02}, the FRB estimate of the $t$th quantile of $[\hat\bth_n]_l$ remains bounded as far as $\hat\bth_{n}$ in equation \eqref{frb1} is a reliable estimate of $\bth$ and more than $(1-t)\%$ of the bootstrap samples contain at least $p$ good (i.e., non-outlying) data points. This means that in FRB, higher quantiles are more robust than the lower ones. Here we show that in a BLFRB bag the former condition guarantees the latter.
\begin{theo}
Let $\check\X = \{(\check{y}_1,\check\z_1^\top)^\top, \ldots, (\check{y}_b,\check\z_b^\top)^\top\}$, be a subsample of size $\dimb = \{\left\lfloor n^\gamma\right\rfloor| \gamma\in\left[0.6,0.9\right]\}$ randomly drawn from $\X$ following the linear model \eqref{lrm}. Assume that the explanatory variables $\check\z_1^\top, \ldots, \check\z_b^\top \in \R^p$ are in general position. 
Let $\hat\bth_{n,b}$ be an MM-estimator of $\bth$ based on $\check\X$ and let $\delta_b$ be the finite sample breakdown point of $\hat\bth_{n,b}$. Then in the BLFRB bag formed by $\check\X$, all the estimated quantiles $\hat{q}_t^{*}$, $t \in (0,1)$ have the same breakdown point equal to $\delta_b$. See the proof in the Appendix.
\label{blfrbrob}
\end{theo}
Theorem \ref{blfrbrob} implies that in the BLFRB setting, lower quantiles are as robust as higher ones with breakdown point equal to $\delta_b$ which can be set close to 0.5. This provides the maximum possible statistical robustness for the quantile estimates. In the proof we show that if $\hat\bth_{n,b}$ is a reliable MM-estimate of $\bth$, then all the bootstrap samples of size $n$ drawn from $\check\X$ are constrained to have at least $p$ good data points.  

\textbf{Table 1} illustrates the upper breakdown points of the BLFRB estimates of quantiles for various dimensions of data and different subsample sizes. The MM-regression estimator is tuned into $50\%$ breakdown point and $95\%$ efficiency at the central model. The results reveal that BLFRB is significantly more robust than the original BLB with LS estimator. Another important outcome of the table is that, when choosing the size of subsamples $b = n^\gamma$, the dimension $p$ of the data should be taken into account; For example for a data set of size $n = 50000$ and $p = 200$, setting $\gamma = 0.6$ or $0.7$ are not the right choices.   
\begin{table}[!t]
\renewcommand{\arraystretch}{1.3}
\caption{Upper breakdown point of the BLFRB estimates of quantiles for MM-regression estimator with $50\%$ breakdown point and $95\%$ efficiency at the Gaussian model.}
\label{table2}
\centering
		\begin{tabular}{c  l  |c  c  c }
				p & \multicolumn{1}{ c| }{n} & $\gamma = 0.6$ & $\gamma = 0.7$ & $\gamma = 0.8$ \\
				\hline
				\multirow{3}{*}{50} 	&	50000		&	 0.425	&	0.475 & 0.491 \\
															&	200000	&	 0.467	&	0.490 & 0.497 \\
															&	1000000	&	 0.488	&	0.497 & 0.499 \\
				\hline															
				\multirow{3}{*}{100} 	&	50000		&	 0.349	&	0.449 & 0.483 \\
															&	200000	&	 0.434	&	0.481 & 0.494 \\
															&	1000000	&	 0.475	&	0.494 & 0.498 \\
				\hline			
				\multirow{2}{*}{200} 
															&	50000		&	 0.197 	&	0.398 & 0.465 \\
															&	200000	&	 0.368	&	0.461 & 0.488	\\
															&	1000000	&	 0.450	&	0.487 & 0.497 \\																																
	      \hline

		\end{tabular}
\end{table}


\section{Numerical Examples}
\label{sec:exp}
In this section the performance of the BLFRB method is assessed by simulation studies. We also perform the simulations with the original BLB method for comparison purposes.
\subsection{Simulation studies}
We generate a simulated data set $\X = \{(y_1,\z_1^\top)^\top, \ldots, (y_n,\z_n^\top)^\top\}$ of size $n=50000$ following the linear  model $y_i = \z_i^\top\bth + \sig_0 e_i$, $(i = 1, \ldots, n)$, where the explaining variables $\z_i$ are generated from \textit{p}-variate normal distribution $\norm_p(\textbf{0},\I_p)$ with $p = 50$, \textit{p}-dimensional parameter vector $\bth = \textbf{1}_p$, noise terms are i.i.d. from the standard normal distribution and noise variance is $\sigma_0^2=0.1$. 

The MM-estimator in the BLFRB scheme is tuned to have efficiency $\mathcal{O} = 95\%$ and breakdown point $BP = 50\%$. The original BLB scheme in \cite{BLB} uses LS-estimator for computation of the bootstrap estimates of $\bth$. 

\begin{figure}[!t]
	  \centerline{\includegraphics[width=8.5cm,height=4.8cm]{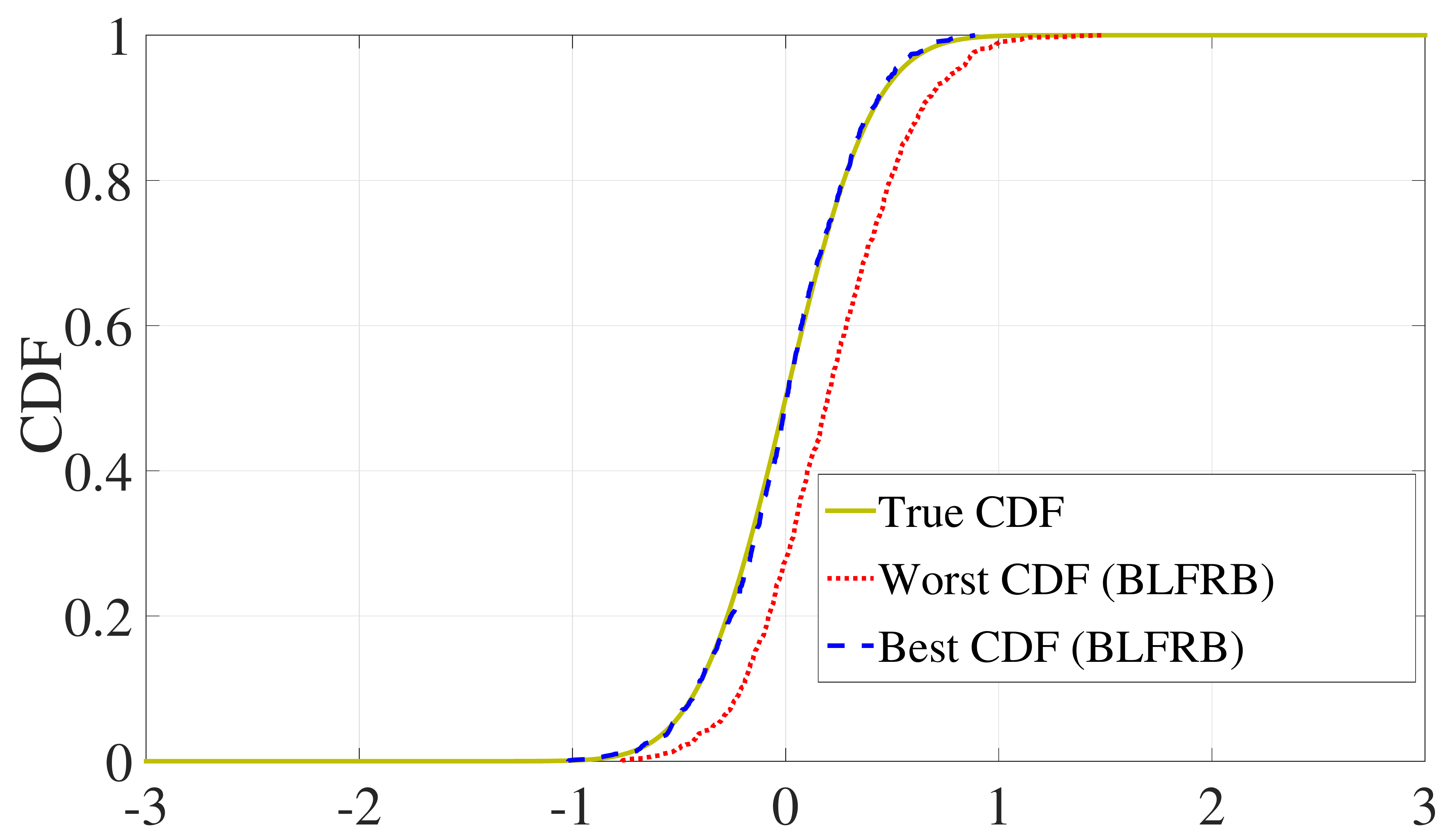}}
	\caption{The true distribution of the right hand side of \eqref{eq:cons} along with the obtained empirical distributions of the left hand side for two elements of $\thf$ with the best and the worst estimates.} 
	\label{consist}
\end{figure}

\begin{figure}[!t]
	  \centerline{\includegraphics[width=8.5cm,height=4.8cm]{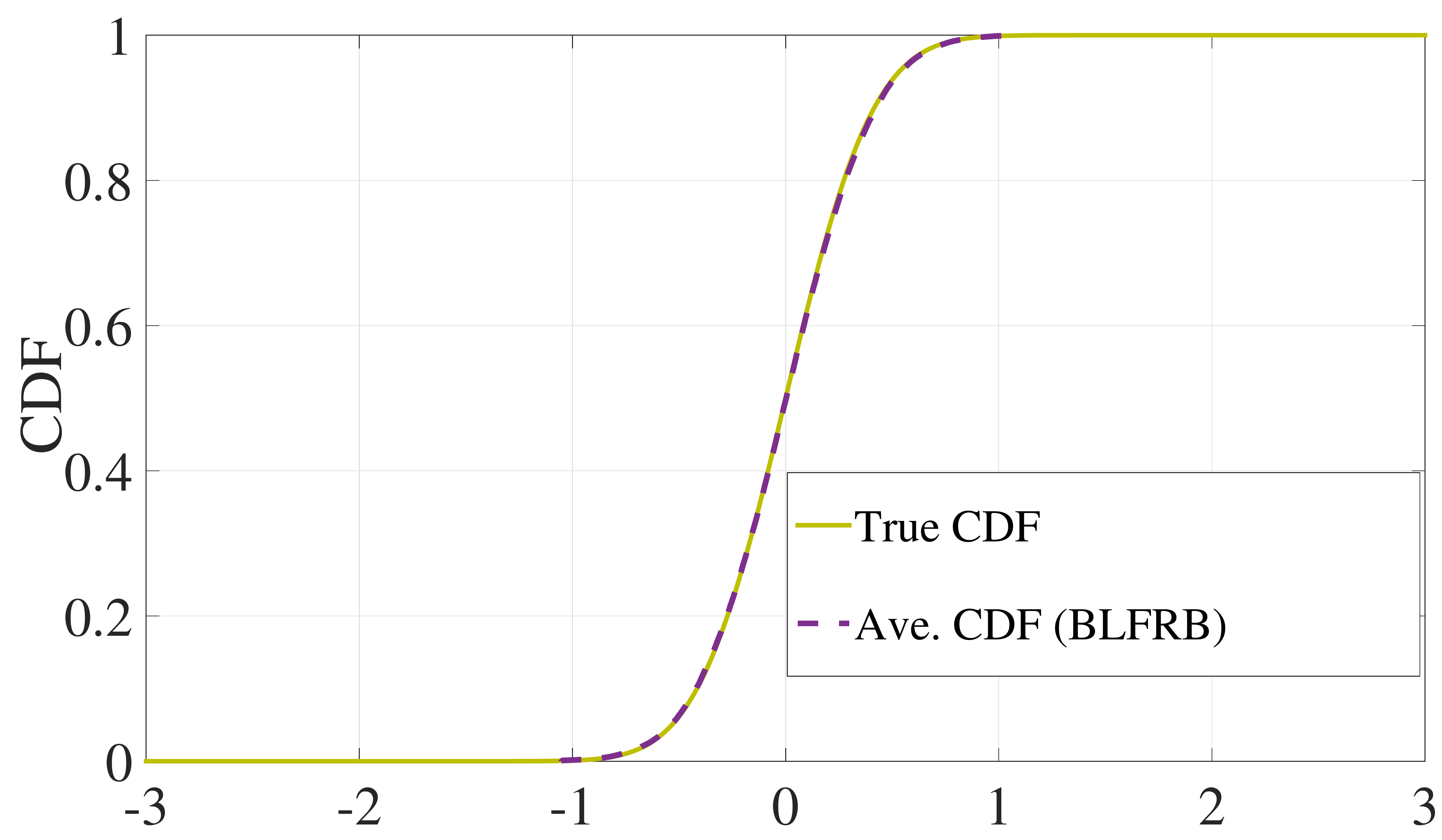}}
	\caption{The average of all $p$ BLFRB estimated distributions, along with the true distribution. Note that the averaged empirical distribution converge to the true cdf. This confirms the results of theorem \ref{cons}.} 
	\label{consist_ave}
\end{figure}

Here, we first verify the result of theorem \ref{cons} in simulation by comparing the distribution of the left hand side of \eqref{eq:cons} with the right hand side. Given the above settings, the right hand side of \eqref{eq:cons} follows $\norm_p(\textbf{0}, \sig_0^2/\mathcal{O}\I_p)$ in distribution \cite[theorem 4.1]{MMest}. We form the distribution of the left hand side,
by drawing a random subsample $\check\X$ of size $b =\left\lfloor 50000^{0.7}\right\rfloor = 1946$ and performing steps 2 and 3 of the BLFRB procedure (i.e., \textbf{Algorithm \ref{alg:BLFRB}}) for $\check\X$ using $r = 1000$ bootstrap samples. 
\textbf{Fig.\ref{consist}} shows the true distribution of $\sqrt{n}(\tho - \bth)$ along with the obtained empirical distributions of $\sqrt{n}(\thf - \hat\bth_{n,b})$ for two elements of $\thf$ with the best and the worst outcomes.
The result of averaging all the $p$ empirical distributions is illustrated in \textbf{Fig.\ref{consist_ave}}, along with the true distribution. Note that the results are in conformity with theorem \ref{cons}. 

\begin{figure}[!t]
	  \centerline{\includegraphics[width=8.5cm,height=4.8cm]{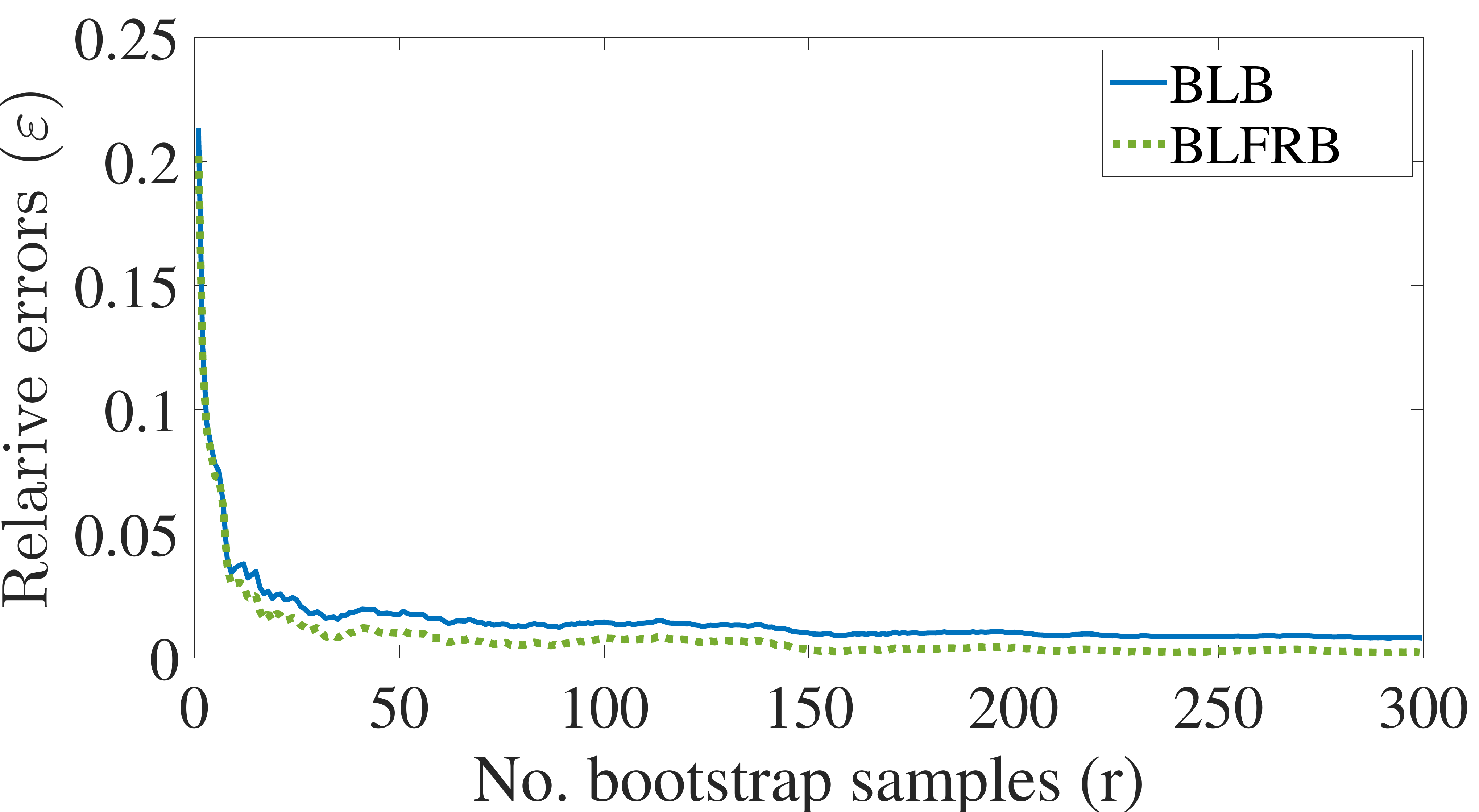}}
	\caption{Relative errors of the BLB (dashed line) and BLFRB (solid line) methods w.r.t. the number of bootstrap samples $r$ are illustrated. Both methods perform equally well when there are no outliers in the data.} 
	\label{50000_50r}
\end{figure}

Next, we compare the performance of the BLB and BLFRB methods. We compute bootstrap estimate of standard deviation (SD) of $\hat\bth_n$ by the two methods. 
In other words, 
 the estimate of uncertainty in step 4 of the procedure (i.e., see \textbf{Fig.\ref{blb}}) for bag \textit{k} is as follows:
\[
\hat\xi_l^{*(k)} = \widehat{\mathrm{SD}}([\ths]_l)=\Bigg(\sum_{j=1}^r{\frac{\Big([\hat\bth^{*(kj)}_{n,b}]_l - [\hat\bth^{*(k\cdot)}_{n,b}]_l\Big)^2}{r-1}}\Bigg)^{1/2},
\]
where 
$[\hat\bth_{n,b}]_l$ denotes the $l$th element of $\hat\bth_{n,b}$ and  
$[\hat\bth^{*(k\cdot)}_{n,b}]_l = \frac{1}{r}\sum_{j=1}^r[\hat\bth^{*(kj)}_{n,b}]_l$.
The step 5 of the procedure for the $l$th element of $\hat\bth_{n,b}$ is obtained by:
\[
\hat\xi_l^{*} = \widehat{\mathrm{SD}}([\hat\bth_{n}]_l)=\frac{1}{s}\sum_{k=1}^s \widehat{\mathrm{SD}}([\ths]_l), \quad l = 1,\ldots,p.
\]
The performance of the BLB and BLFRB are assessed by computing a relative error defined as:
\[ 
\eps = \frac{\left|\widehat{\mathrm{SD}}(\hat\bth_n) - \overline{\mathrm{SD}}_o(\hat\bth_n)\right|}{\overline{\mathrm{SD}}_o(\hat\bth_n)},
\]
where $\widehat{\mathrm{SD}}(\hat\bth_n) = \frac{1}{p} \sum_{l=1}^p \widehat{\mathrm{SD}}([\hat\bth_{n}]_l)$ and $\overline{\mathrm{SD}}_o(\hat\bth_n) = \sig_0/\sqrt{n\mathcal{O}}$ is (approximation of) the average  standard deviation of $\hat\bth_n$  based on the asymptotic covariance matrix \cite{MMest} (i.e., $\mathcal{O}$ is 0.95 for the MM-estimator and 1 for
the LS-estimator).
The bootstrap setup is as follows; 
Number of disjoint subsamples is  $s = 25$, size of each subsample is $b =\left\lfloor n^\gamma\right\rfloor = 1946$ with $\gamma = 0.7$, maximum number of bootstrap samples in each subsample module is $r_{max} = 300$. 
We start from $r = 2$ and continually add a new set of bootstrap samples (while $r<r_{max}$) to subsample modules. The convergence of relative errors w.r.t. the number of bootstrap samples $r$ are illustrated in \textbf{Fig.\ref{50000_50r}}. Note that when the data is not contaminated by outliers, both methods perform similarly in terms of achieving lower level of relative errors for higher number of bootstrap samples.

\begin{figure}%
\centering
\begin{mdframed}
\subfloat[][]{\label{50000_50rob_blb}\includegraphics[width=8.3cm,height=4.8cm]{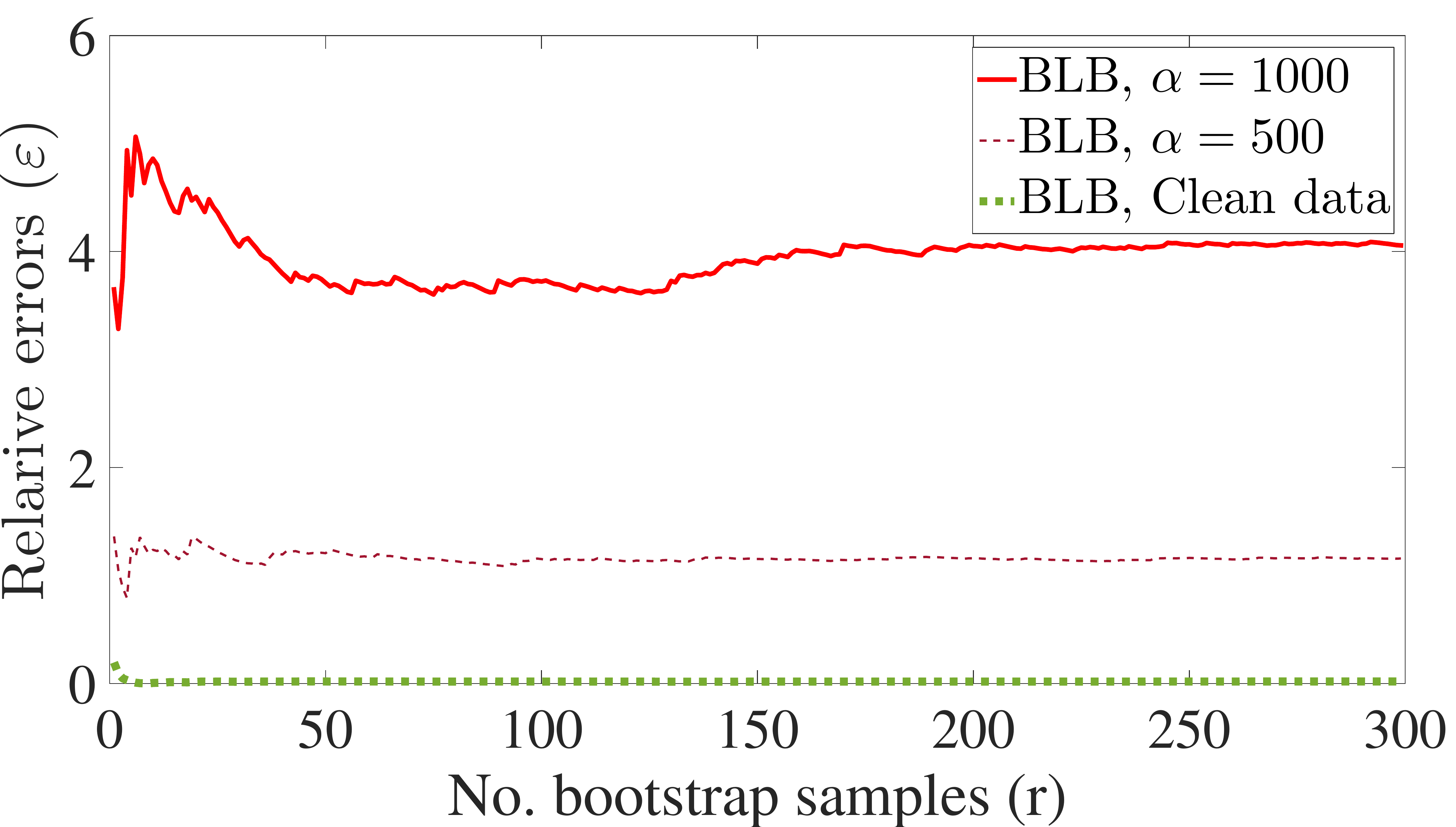}}%
\qquad
\subfloat[][]{\label{50000_50rob_blfrb}\includegraphics[width=8.3cm,height=4.8cm]{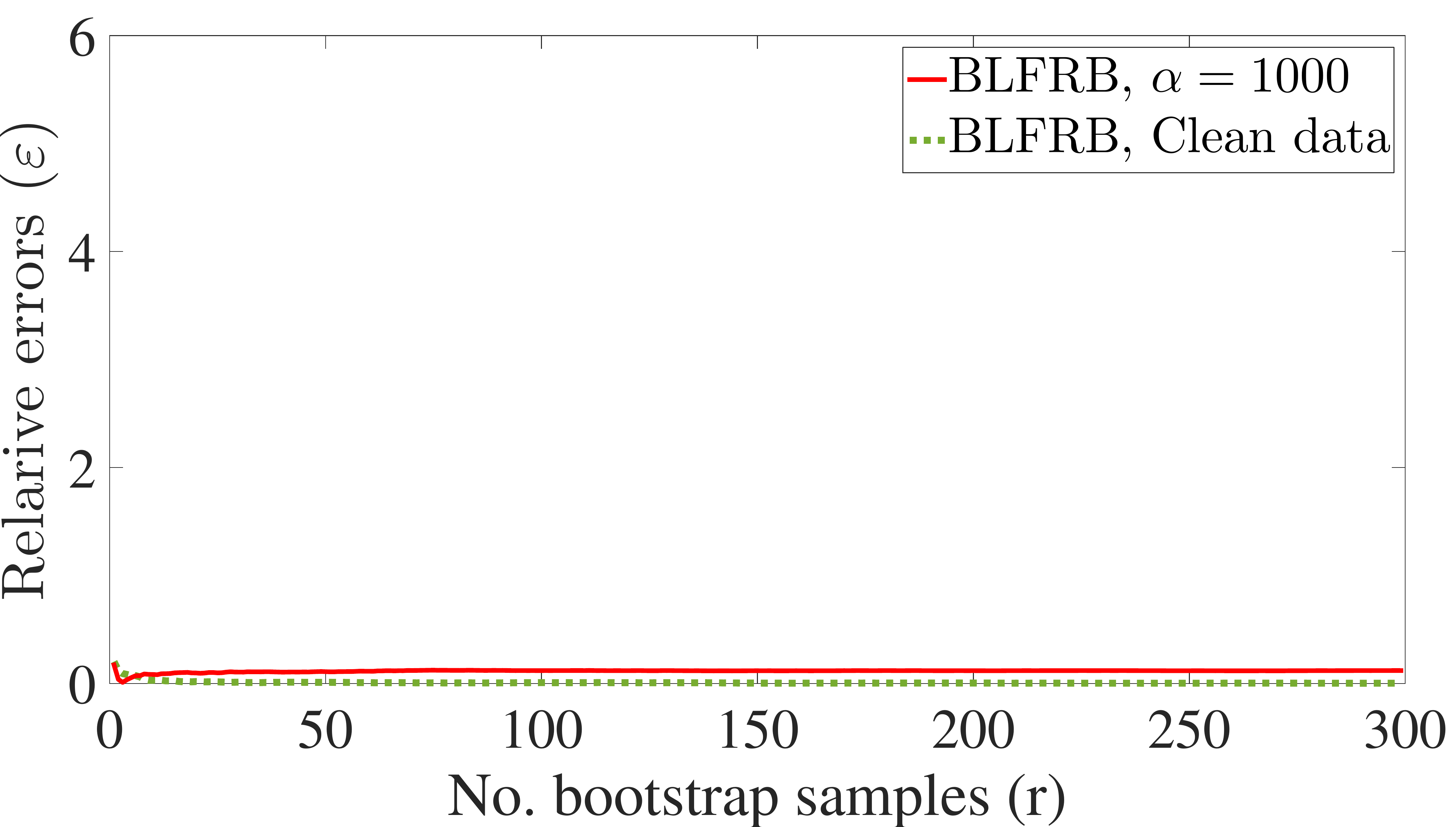}}
\end{mdframed}
\caption{\textbf{(a)} Relative errors of the BLB method illustrating severe lack of robustness in face of only one outlying data point. \textbf{(b)} Relative errors of the BLFRB method illustrating the reliable performance of the method in the face of severely contaminated data.}%
\label{fig:test}	
\end{figure}

We study the robustness properties of the methods using the above settings. According to theorem \ref{blbrob}, only one outlying data point is sufficient to drive the BLB estimates of $\widehat{\mathrm{SD}}(\hat\bth_n)$ over any finite bound. To introduce such outlier, we randomly choose one of the original data points and multiply it by a large number $\al$ . Such contamination scenario resembles misplacement of the decimal point in real world data sets. Lack of robustness of the BLB method is illustrated in \textbf{Fig.\ref{50000_50rob_blb}} for $\al = 500$ and $\al = 1000$. 

According to \textbf{Table 1}, for the settings of our example the upper breakdown point of BLFRB quantile estimates is $\delta_b = 0.475$. Let us asses the statistical robustness of the BLFRB scheme by severely contaminating the original data points of the first bag. We multiply $40\%$ ($\left\lfloor 0.4\times b \right\rfloor = 778$) of the data points  by $\al = 1000$. As shown in \textbf{Fig.\ref{50000_50rob_blfrb}}, BLFRB still performs highly robust despite such proportion of outlying data points.   

\begin{figure}[!t]
	  \centerline{\includegraphics[width=8.5cm,height=4.8cm]{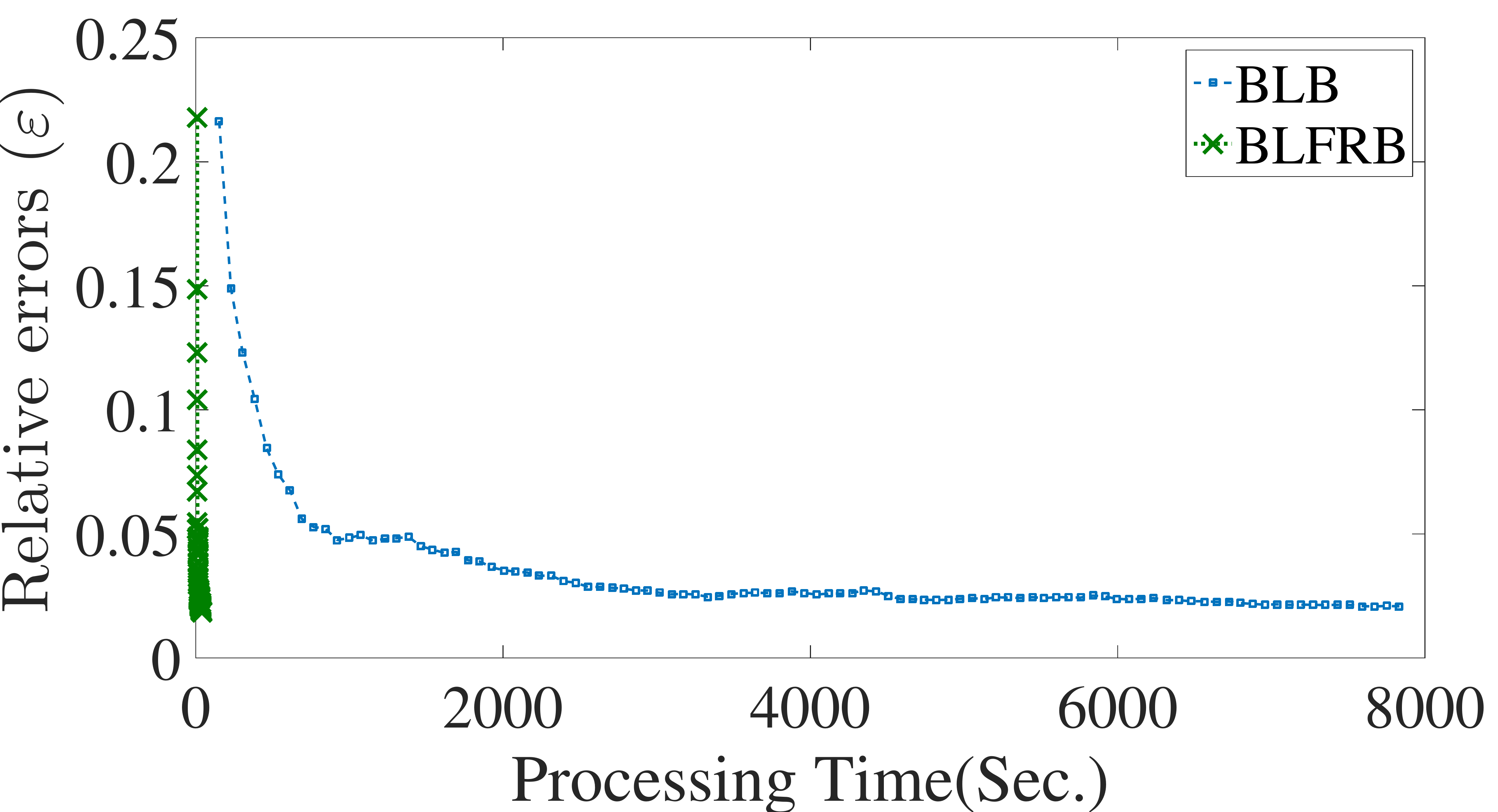}}
	\caption{Relative errors $\eps$ w.r.t. the required processing time of each BLB and BLFRB iteration. The BLFRB is significantly faster to compute as the (re)computation of the estimating equations is not needed in this method.} 
	\label{50000_50t}
\end{figure}

Now, let us make an intuitive comparison between computational complexity of the BLB and BLFRB methods by using the MM-estimator in both methods. We use an identical computing system to compute bootstrap standard deviation (SD) of $\hat\bth_n$ by the two methods. The computed $\eps$ and the cumulative processing time are stored after each iteration (i.e.,adding new set of bootstrap samples to the bags). 
\textbf{Fig.\ref{50000_50t}}, reports relative errors w.r.t. the required cumulative processing time after each iteration of the algorithms. The BLFRB is remarkably faster since the method avoids solving estimating equations for each bootstrap sample. 
 \subsection{Real world data}
Finally, we use the BLFRB method for bootstrap analysis of a real large data set. We consider the simplified version of the the Million Song Dataset (MSD) \cite{MSD}, available on the UCI Machine Learning Repository \cite{UCI}. The data set $\X = \{(y_1,\z_1^\top)^\top, \ldots, (y_n,\z_n^\top)^\top\}$ contains $n = 515345$ music tracks, where $y_i$ (i.e., $i = 1, \ldots, n)$ represents the released year of the $i$th song (i.e., ranging from 1922 to 2011) and $\z_i \in \R^p$ is a vector of $p = 90$ different audio features of each song. The used features are the average and non-redundant covariance values of the timbre vectors of the song.

\begin{figure}[!t]
	  \centerline{\includegraphics[width=8.5cm,height=4.8cm]{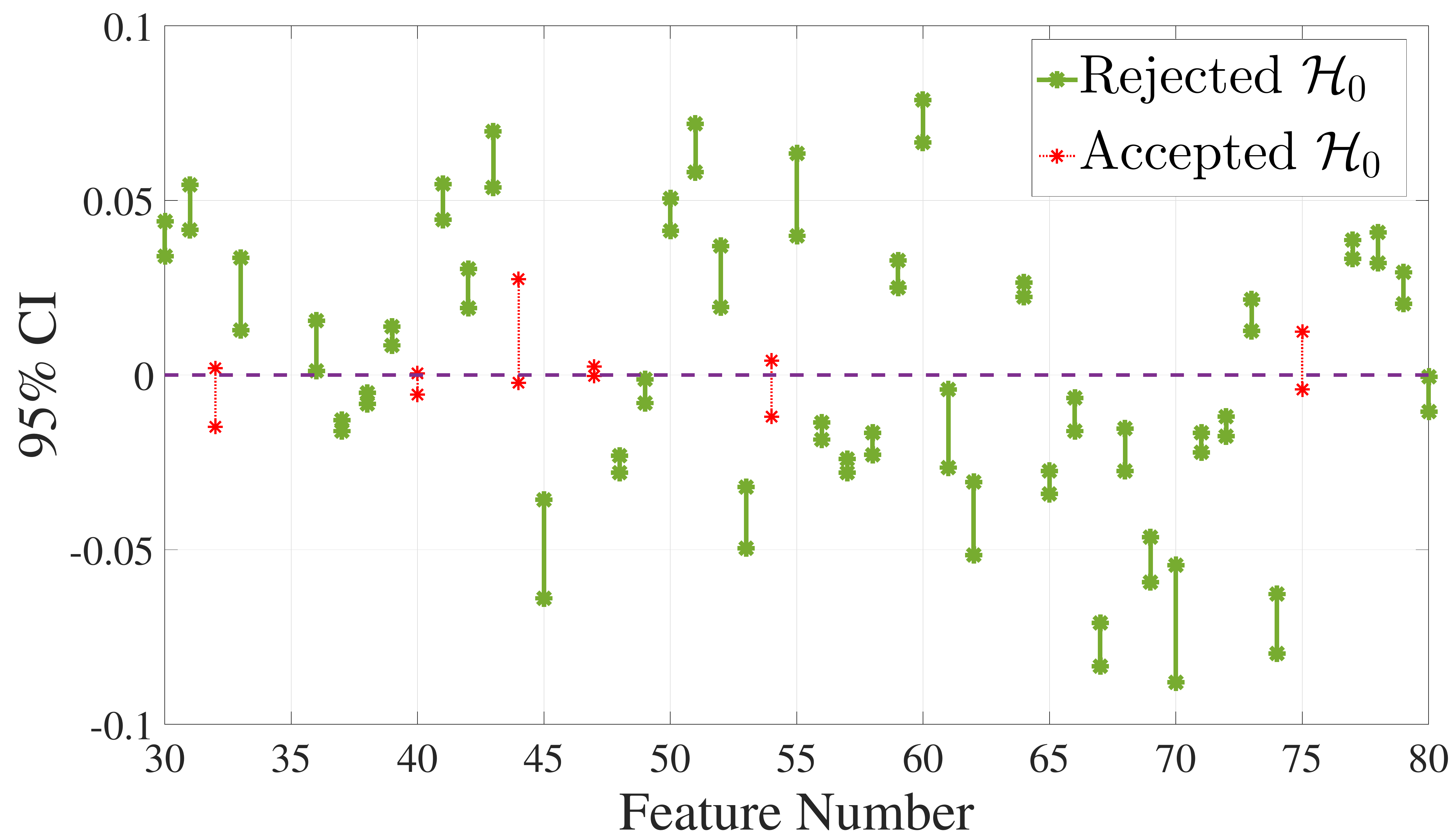}}
	\caption{The $95\%$ confidence intervals computed by BLFRB method is shown for some of the audio features of the MSD data set. The null hypothesis in accepted for those features having 0 inside the interval.} 
	\label{MSDCI}
\end{figure}

The linear regression can be used to predict the released year of a song based on its audio features. We use the BLFRB method to conduct a fast, robust and scalable bootstrap test on the regression coefficients. 
In other words, considering the linear model $y_i = \z_i^\top\bth + \sig_0 e_i$, we use BLFRB for testing hypothesis $\h_0: \left\lfloor \bth \right\rfloor_l = 0$ vs. $\h_1: \left\lfloor \bth \right\rfloor_l \neq 0$, where $\left\lfloor \bth \right\rfloor_l$ (i.e., $l = 1, \ldots, p$) denotes the $l$th element of $\bth$. The BLFRB test of level $\alpha$ rejects the null hypothesis if the computed $100(1 - \alpha)\%$ confidence interval does not contain $0$. 
Here we run the BLFRB hypothesis test of level $\alpha = 0.05$ with the following bootstrap setup; 
Number of disjoint subsamples is  $s = 51$, size of each subsample is $b =\left\lfloor n^\gamma\right\rfloor = 9964$ with $\gamma = 0.7$, number of bootstrap samples in each subsample module is $r = 500$. 
Among all the 90 features, the null hypothesis is accepted only for 6 features numbered:  $32, 40, 44, 47, 54, 75$. \textbf{Fig.\ref{MSDCI}} shows the computed $95\%$ CIs of the features. In order to provide a closer view, we have only shown the results for feature numbers 30 to 80. These results can be exploited to reduce the dimension of the data by excluding the ineffective variables from the regression analysis.  
  
Using the BLFRB method with the highly robust MM-estimator, we ensure that the computational process is done in a reasonable time frame and the results are not affected by possible outliers in the data. Such desirable properties are not offered by the other methods considered in our comparisons.

\section{Conclusion}\label{sec:conc}
In this paper, a new robust, scalable and low complexity bootstrap method is introduced with the aim of finding parameter estimates and confidence measures for very large scale data sets. The statistical properties of the method including convergence and robustness are established using analytical methods. While the proposed BLFRB method is fully scalable and compatible with distributed computing systems, it is remarkably faster and significantly more robust than the original BLB method \cite{BLB}. 


%

\appendices
\section{}
\label{sec:appa}
Here we provide the proofs of the theoretical results of sections \ref{subsec:Consis} and \ref{subsec:MMest}. 

\textbf{Proof of theorem \ref{cons}}: Given that $\f$ is Donsker class, as $n \rightarrow \infty$: 
\beq\G_n = \sqrt{n}(\Po - P) \stackrel{d}{\rightarrow} \G_p,\eeq
where $\G_p$ is the \emph{P-Brownian bridge process} and the notation $\stackrel{d}{\rightarrow}$ denotes convergence in distribution. 
According to \cite[theorem 3.6.3]{vv96} as $b,n \rightarrow \infty$:
\beq\Gb = \sqrt{n}(\Pb - \Ps) \stackrel{d}{\rightarrow} \G_p.\eeq
Thus, $\G_n$ and $\Gb$ converge in distribution to the same limit.
The functional delta method for bootstrap \cite[theorem 23.9]{vv98} in conjunction with \cite[lemma 1]{BLBsup} imply that, for every Hadamard-differentiable function $\phi$:
\beq \sqrt{n}(\phi(\Po) - \phi(P)) \stackrel{d}{\rightarrow} \phi'_{P}(\G_p), \eeq
and conditionally on $\Ps$,  
\beq \sqrt{n}(\phi(\Pb) - \phi(\Ps)) \stackrel{d}{\rightarrow} \phi'_{P}(\G_p), \eeq
where $\phi'_{P}(\G_p)$ is the derivative of $\phi$ w.r.t. $P$ at $\G_p$. 
Thus, conditionally on $\Ps$, 
\beq\sqrt{n}(\thb - \ths) \stackrel{d}{=} \sqrt{n}(\tho - \bth).\label{convblb}\eeq

According to \cite[equation 5 and 6]{FRB08} and given that $\tho$ can be expressed as a solution to a system of smooth FP equations:
\beq\sqrt{n}(\thf - \ths) \stackrel{d}{=} \sqrt{n}(\thb - \ths).\label{convfrb}\eeq 
Form \eqref{convblb} and \eqref{convfrb}: 
\beq\sqrt{n}(\thf - \ths) \stackrel{d}{=} \sqrt{n}(\tho - \bth)\label{convblfrb}\eeq 
which concludes the proof. $\square$

The following lemma is needed to prove theorems \ref{blbrob} and \ref{blfrbrob}.
\begin{lem}
Let $\check\X = (\check\x_{1} \ \cdots \  \check\x_{\dimb})$ be a subset of size $\dimb = \{\left\lfloor n^\gamma\right\rfloor| \gamma\in\left[0.6,0.9\right]\}$ randomly resampled without replacement from a big data set $\X$ of size $n$. Let $\X^*$ be a bootstrap sample of size $n$ randomly resampled with replacement from $\check\X$ (i.e., or equivalently formed by assigning a random weight vector $\n^{*} = (n_{1}, \ldots, n_{b}^*)$ from $Multinomial(n, (1/\dimb)\textbf{1}_{\dimb})$ to columns of $\check\X$\;). Then:
\[
\lim_{n \to \infty} Pr\{\check\x_{(i)} \notin \X^* | \text{ for any } \check\x_{(i)} \in \check\X\} \to 0.
\]    
\label{lem1}
\end{lem}

\textbf{Proof of lemma \ref{lem1}}: Consider an arbitrary data point $\check\x_{(i)} \in \check\X$. The probability that $\check\x_{(i)}$ does not occur in a bootstrap sample of size $n \to \infty$ is:
\begin{gather*}
\lim_{n \to \infty} Pr(Binomial(n,\frac{1}{n^\gamma}) < 1) =\\ \lim_{n \to \infty}(1 - \frac{1}{n^\gamma})^n
= \lim_{n \to \infty} \exp \Big(\frac{\ln (1 - \frac{1}{n^\gamma})}{1/n}\Big) = \\ \lim_{n \to \infty} \exp \Big(-\frac{\gamma n^2}{n^{\gamma+1}-n}\Big) = 0. \quad\square
\end{gather*}
Such probability for $n = 20000$ and $\gamma = 0.7$ is $3.3 \times 10^{-9}$. 

\textbf{Proof of theorem \ref{blbrob}}: Let $\check\x_{(i)} \in \check\X $ be an outlying data point in $\check\X$. According to lemma \ref{lem1}, all bootstrap samples drawn from that subsample will be contaminated by $\check\x_{(i)}$. This is sufficient to break all the LS replicas of the estimator in that bag and consequently ruining the end result of the whole scheme. $\square$ 

\textbf{Proof of theorem \ref{blfrbrob}}: 

According to \cite[Theorem 2]{FRB02}, The FRB estimate of $q_t$ remains bounded as far as: 
\begin{itemize}
\item[1.] $\hat\bth_{n}$ in equation \eqref{frb1} is a reliable estimate of $\bth$, and
\item[2.] More than $(1-t)\%$ of the bootstrap samples contain at least $p$ "good" (i.e., non-outlying) data points.
\end{itemize}
The first condition implies that, in a BLFRB bag if $\hat\bth_{n,b}$ is a corrupted estimate then all bootstrap estimates $\hat{q}_t^{*}$, $t \in (0,1)$ will break as well. In the rest of the proof we show that if the percentage of outliers in $\check\X$ is such that $\hat\bth_{n,b}$ is still a reliable estimate of $\bth$, then all the bootstrap samples drawn from $\check\X$ contain at least $p$ good (non-outlying) data points. This suffices for all $\hat{q}_t^{*}$, $t \in (0,1)$ to remain bounded.   

Let $\hat\bth_{n,b}$ be an MM-estimate of $\bth$. Let the initial scale of the MM-estimator obtain by a high breakdown point S-estimator. The finite sample breakdown point of the S-estimator for a subsample of size $b$ is as follows: 
\[
\delta_b^S = \frac{\left\lfloor b/2\right\rfloor - p + 2}{b},
\]  \cite[Theorem 8]{FRB87}. Given that $\hat\bth_{n,b}$ is a reliable estimate, the initial S-estimate of the scale parameter is not broken. This implies that there exist at least $h = b - \left\lfloor b/2\right\rfloor + p - 1$ good data points in general position in $\check\X$. It is easy to see that $h > p$. Applying Lemma 1 for each of the good points concludes that the probability of drawing a bootstrap sample of size $n$ with less than $p$ good data points goes to zero for large $n$, which is the case of big data sets. $\square$

\ifCLASSOPTIONcaptionsoff
  \newpage
\fi

\bibliography{BLFRB_Journal_v1}
\bibliographystyle{IEEEtr}

\end{document}